\documentclass[10pt,draft,notref,notcite]{article}

\usepackage{color,longtable}
\usepackage{amssymb,graphicx}

\newcommand{\be}{\begin{eqnarray}}
\newcommand{\ee}{\end{eqnarray}}

\usepackage{amsfonts,amsmath}
\usepackage{latexsym}

\def\cG{{\cal{G}}}
\def\cH{{\cal H}}
\def\cN{{\cal N}}

\def\rd{{\rm{d}}}

\def\a{{\alpha}}
\def\b{{\beta}}

\def\E{{\rm E}_{10}}
\def\K{{\rm K}(\E)}

\def\lae{{\mathfrak{e}_{10}}}
\def\lag{{\mathfrak{g}}}

\def\las{{\mathfrak{sl}_{2}}}

\def\ni{\noindent}

\begin{document}

\begin{center}
{\bf \Large Complexity and the Big Bang}\\[7mm]
  Hermann Nicolai\\[4mm]
{\sl  Max-Planck-Institut f\"ur Gravitationsphysik\\
     Albert-Einstein-Institut \\
     M\"uhlenberg 1, D-14476 Potsdam, Germany\\[1mm]
 Email: {\tt nicolai@aei.mpg.de}} \\[17mm]
\begin{minipage}{12cm}\footnotesize
\textbf{Abstract:} After a brief review of current scenarios for the resolution 
and/or avoidance of the Big Bang, an alternative hypothesis is put forward 
implying an infinite increase in complexity towards the initial singularity.
This may result in an effective non-calculability 
which would present an obstruction to actually reaching the beginning 
of time.  This proposal is motivated by the appearance of certain infinite-dimensional 
duality symmetries of indefinite Kac--Moody type in attempts to 
unify gravity with the fundamental matter interactions,
and deeply rooted in properties of Einstein's theory.\\

\end{minipage}
\end{center}

\vspace{6cm}

\newpage

\noindent

{\noindent\bf 1. Introduction.}
The beginning of the Universe is shrouded in mystery. Observations indicate 
that it came into being with a Big Bang, where the space-time geometry
and physical quantities characterizing matter (density, pressure, temperature, 
{\em etc.}) appear to become singular,
see {\em e.g.} \cite{KT,M} for an introduction. It seems generally agreed that the resolution 
of these singularities will require the reconciliation of general relativity and 
quantum theory into a consistent theory of quantum gravity. Addressing this 
challenge should pave the way for a proper understanding of what precisely 
`happened' at the cosmic instant $\tau =0$, and answer the question whether the 
Big Bang represents a true beginning, or whether the history of our universe 
extends into a possibly infinite past. 

In this Note I wish to outline an approach to the problem of the initial singularity 
that is motivated by the magic link between Einstein's theory and its (supersymmetric) 
generalizations on the one hand, and certain infinite-dimensional duality 
symmetries, and the hyperbolic Kac--Moody algebra $\lae$, in particular, on the other.
The main observation is that properties of these algebras may introduce a  
mathematically well defined element of non-calculability
which may forever shield the initial singularity from resolution. 

\vspace*{0.2cm}
\ni{\bf 2. Brief survey of previous ans\"atze.} 
Existing approaches to quantum gravity propose specific ways to deal with
the initial singularity, thereby also offering potential answers to the questions
posed above: 

\vspace*{0.1cm}
\ni{\em Creation `from nothing'.} The idea that the universe was created by quantum tunneling 
from nothing, and without an initial singularity, was put forward already some time ago 
\cite{BEG,Vilenkin1} \footnote{An even earlier variant of this idea is B. DeWitt's suggestion
 that the wave function of the universe might have vanishing support
 on singular 3-geometries \cite{DeWitt}; see also \cite{KKN,Kiefer0} for
 more recent results in this direction. } and motivates 
 the celebrated Hartle-Hawking ansatz for the wave function of the universe \cite{HH}.
According to the latter, Lorentzian space-time can be explained as emerging 
from a Euclidean instanton. The switch from Lorentzian to Euclidean signature 
enables one to `round off' the past before the Big Bang into a cap, 
such that the beginning of time can no longer be located, and the singularity 
simply `disappears'. However, the precise meaning of the term `nothing' 
here raises subtle ontological issues that reach beyond physics, see {\em e.g.} \cite{Tang}: 
for instance, is this a creation out of {\em literally} nothing, or is it a process to 
be embedded in a larger entity (a megaverse)?

\vspace*{0.1cm}
\ni{\em Quantum bounce.}
Canonical approaches to quantum gravity posit that singularity resolution can
be achieved by replacing the classical time evolution of the spatial geometry 
by the quantum mechanical evolution of  a wave function(al)
smearing the 3-geometries over the singular classical trajectories 
\cite{Kiefer,Bojo,Ashtekar} (for possible realizations of singularity 
avoidance in the framework of covariant loop quantum gravity, see \cite{RV}).
The bang would be replaced by a quantum bounce, and preceded by a 
collapsing universe, and would therefore not represent a beginning of time. 
The ancestor universe itself must either have originated from an infinite past, 
or be part of a sequence of repeated bounces. Cosmic bounces
are also central to the ekpyrotic universe \cite{Turok}.

\vspace*{0.1cm}
\ni{\em Avoiding the singular point.} A fundamental theory may
admit an operation allowing for the cosmic scale factor $a$ to `jump over' 
the singular point, for instance if one can analytically continue its dynamics 
into the complex $a$-plane. Pre-Big Bang string cosmology relies on the 
purported UV completeness of string theory to argue that such an operation can 
be safely implemented, {\em e.g.}  by means of a $T$ duality transformation 
linking different `phases' of space-time \cite{GV}.
Likewise, but with very different reasoning, Penrose's conformal cyclic cosmology 
postulates a `reset' $a \!=\! \infty \rightarrow a'\!\equiv\, a^{-1}\!=\!0$ 
at the end of each `\ae on' \cite{Penrose1}, a process whose dynamical origin
remains unclear, however.

\vspace*{0.1cm}
\ni{\em Eternal inflation.} Many variants of inflationary cosmology \cite{MRV}
are based on the hypothesis that baby universes can arise by quantum mechanical 
bubble nucleation, a process that is assumed to take place continually.
In this picture, a megaverse, of which our universe is merely an infinitesimal part,
would simply  `exist' with neither beginning nor end, possibly also allowing for other 
kinds of universes, as suggested by the string landscape \cite{Sch}.
However, it has been shown that there then always exist geodesically incomplete
past directed curves \cite{AGV}. This excludes the possibility that our universe 
can be embedded in a  megaverse space-time geometry that is completely 
free of singularities (but see \cite{Ellis}). For the resolution of 
multiple past and future singularities linked to the
appearance and disappearance of baby universes this scenario
has to appeal to other ideas. \\[2mm]
A common feature of these proposals is that they often rely on the assumption
that there is a basic `simplicity' about the universe and its beginning, thus  extrapolating 
Einstein's cosmological principle to quantum  cosmology. This is usually
done by restricting the infinitely many degrees of freedom to a finite number of 
variables (such as the cosmic scale factor and a number of spatially homogeneous 
fields), or alternatively by means of purely heuristic manipulations of an ill-defined
gravitational path integral or Wheeler-DeWitt operator. Contrary
to these often stated views, I here want to argue that there is no such
`simplicity' about the beginning of the universe, in the sense that
the underlying fundamental theory could exhibit an unbounded complexity towards
the initial singularity that may turn out to be an inherent and ineluctable 
feature of quantum gravity.

\vspace*{0.2cm}
\ni{\bf 3. Reformulating  gravitational dynamics near the singularity.} 
Before getting to the main point (in section 5) we first need some preparation.
We start from the space-time metric (setting the shift to zero)
\be
\rd s^2 = - (N \rd x^0)^2 + {\tt g}_{mn} \rd x^m \rd x^n
\qquad (m,n,...= 1,...,d)
\ee
where $d$ is the number of spatial dimensions.
Denoting the momenta canonically conjugate to the spatial metric components 
${\tt g}_{mn}$  by $\Pi^{mn}$,  the Hamiltonian
constraint  reads (see {\em e.g.} \cite{Kiefer})
\be\label{H0}
\cH \,=\, G_{mn|pq} ({\tt g}) \,\Pi^{mn} \Pi^{pq} \,-\, 
\sqrt{{\tt g}}   R^{(d)}({\tt g}) \,+ \, \cH_{\rm matter}  \,\approx \, 0 
\ee
Here $G_{mn|pq}({\tt g})$ is the DeWitt metric, which is famously {\em indefinite},
and the matter part of the Hamiltonian $\cH_{\rm matter}$
need not be specified at this point.
For the mathematical analysis of the dynamics near the singularity 
{\em \`a la} Belinski-Khalatnikov-Lifshitz (BKL)  \cite{BKL} 
one singles out the logarithmic
scale factors $\b^a$, parametrizing the off-diagonal metric components  
in Iwasawa form \cite{DaHeNi03}
\be\label{Iwasawa}
{\tt g}_{mn} = \sum_{a=1}^{d} e^{- 2\b^a} \cN^a{}_m \cN^a{}_n 
\ee
where the matrices $\cN^a{}_m$ are upper triangular, with $\cN^a{}_m =1$ for $a=m$.
It is convenient to choose  a gauge for the lapse function $N$ such that, 
with proper time $\tau$, the time coordinate $t\sim - \log \tau$ becomes `Zeno-like'
({\em i.e.} stretching the finite time interval in $0 < \tau < \tau_0$ to infinity)
and the singularity is located in the infinite future ($t=\infty$). A detailed 
analysis now shows that with these variables the Hamiltonian constraint 
(\ref{H0}), at a given spatial point, and with different matter couplings, can 
be rewritten as \cite{DaHeNi03}
\be \label{H1}
\cH(\b^a, \pi_{a},...) \, = \,   
G^{ab} \pi_a \pi_b \, + \, \sum_A c_A(...)  e^{- 2 w_A (\b)} 
\ee
Here, $\pi_a$ are the canonical momenta conjugate to the scale factors $\b^a$, 
and $G_{ab}$ is the Lorentzian metric obtained by restricting the DeWitt 
metric to diagonal metric degrees of freedom. The dots in (\ref{H1}) stand for 
all other canonical degrees of freedom (off-diagonal metric components, spatial gradients, 
matter, {\em etc.}). 

The exponential terms in (\ref{H1}) involve {\em linear forms} 
$w_A(\b) \equiv w_{A a}  \b^a$, where the vectors $w_{A}$ characterize the various 
contributions to the Hamiltonian, and the prefactors $c_A$ are (complicated) functions 
of the canonical variables other than $\b^a$ and $\pi_a$, see \cite{DaHeNi03} for details. 
The walls normal to the vectors $w_{A}$ partition the $\b$-space into chambers.
In the near singularity (BKL) limit the Hamiltonian dynamics simplifies radically,
being dominated by a number of {\em leading wall forms} $\{ w_i \,|\, i=1,...,r \}$, and is
reduced  to that of a relativistic billard ball bouncing off the walls
of a fundamental chamber in $\b$-space, see also \cite{Misner} for an
equivalent description. The occurrence (or not) of chaotic 
metric oscillations depends on the `shape' of this chamber,
where the `shape' itself depends on the matter content and the number of
space-time dimensions: chaotic oscillations happen whenever the 
billiard wedge is located within the light-cone in $\beta$-space \cite{DHJN}.
This description thus unifies previous analyses based on case-by-case
studies \cite{DHHS}.\footnote{ Because the BKL analysis relies on {\em causal decoupling} of 
spatial points, it applies only to {\em space-like} (cosmological) singularities, but not
to time-like or null singularities.}

\vspace*{0.2cm}
\ni{\bf 4. Group theoretical transmutation of the gravitational Hamiltonian.} 
The key fact is now that there exists a detailed, though still very incomplete,  
correspondence of the (matter coupled) gravitational dynamics with a purely 
group theoretical construct \cite{DaHeNi02}. To exhibit
it, one identifies the space of scale factors $\b^a$ 
with the Cartan subalgebra (CSA) of a certain indefinite Kac-Moody algebra $\mathfrak{g}$;
the metric $G_{ab}$ from (\ref{H1}) then becomes the Cartan-Killing metric on the CSA.
Crucially, the indefiniteness of the DeWitt metric implies that the Cartan-Killing metric 
is also indefinite. Identifying the simple roots of $\lag$ with the dominant
wall forms $w_i$, the Cartan matrix defined by \cite{DaHe01}
\be\label{A}
A_{ij} \equiv G^{ab} w_{ia} w_{jb}   \qquad\quad  (1\leq i,j \leq r )
\ee
is likewise indefinite (for simplicity we take $\lag$ to be simply laced).
The dynamics take place on the infinite-dimensional coset space $\cG/K(\cG)$, 
where $\cG$ is the (non-compact) `group' obtained by formally exponentiating $\mathfrak{g}$, and $K(\cG)$ 
its `maximal compact subgroup'. Adopting an Iwasawa gauge analogous to
(\ref{Iwasawa}), it is governed by the Hamiltonian 
\be\label{HKM}
\cH_{KM} (\b^a,\pi_a, ... ) \,=\,  
  G^{ab} \pi_a \pi_b  \, + \, \sum_{\a > 0} \sum_{s=1}^{{\rm mult}(\a)}
      \Pi_{\a ,s}^2(...) e^{-2\a(\b)}  
\ee
where the sum on the r.h.s. ranges over all positive roots $\a$ of $\lag$ and their 
multiplicities (with $\a(\b) \!\equiv\! \a_a\b^a$), and $\Pi_{\a,s}$ parametrize
the non-CSA degrees of freedom. Restricting the dynamics to the 
leading walls and to the simple roots, respectively, one finds exact agreement 
between (\ref{H1}) and (\ref{HKM}) \cite{DaHeNi03}. More generally,
(\ref{HKM})  implies {\em null geodesic motion on} $\cG/K(\cG)$.
The correspondence between the two models can then be shown to extend 
to first order spatial gradients for the gravitational Hamiltonian, 
and to roots $\alpha$ of low height for the coset Hamiltonian (\ref{HKM}) \cite{DaHeNi02}.

As is well known, the Hamiltonian constraint must be imposed at each spatial point.
By contrast, there is no `space' in (\ref{HKM})! Moreover, instead of 
finitely many contributions $c_A$ to the potential as in (\ref{H1}), the sum 
on the r.h.s. of  (\ref{HKM}) has an {\em infinite} number of contributions. 
The conjecture put forward in \cite{DaHeNi02} then amounts to 
saying that the spatial field theory degrees of freedom now `reside' in  the 
infinite-dimensional Lie algebra $\lag$. This picture of the evolution 
as a {\em one-dimensional motion} is in accord with the fact that 
cosmological evolution can be described as geodesic motion of 
a point particle in the moduli space of spatial geometries \cite{G,TW} -- 
except that this moduli space is now replaced
by the coset space $\cG/K(\cG)$. To be sure, numerous 
conceptual and mathematical questions remain. 
Most important, these concern the precise nature of the mechanism by which 
space and space-time\footnote{Since we are working in a Hamiltonian framework, 
`time' would have to emerge from the timeless Wheeler-DeWitt equation, in analogy with 
more standard canonical approaches \cite{Kiefer}.}, as well as concomitant
features such as general covariance and gauge symmetry,
are supposed to emerge
from a purely group theoretical construct, and the key role that quantization
and the incorporation of fermions are expected to play in it 
(but see {\em e.g.} \cite{Ganor,KKN,DS}).

In the remainder I will focus on the most important case, maximal 
supergravity in eleven space-time dimensions (`M theory') \cite{CJS,Ju85}, for which 
$\lag \!=\! \lae$. This is the maximally extended 
(= maximal rank) hyperbolic Kac-Moody algebra,
with `group' $\cG = \E$, which for myriad reasons is a prime candidate symmetry 
for a Planck scale unified theory of quantum gravity (but see \cite{West}).

\vspace*{0.2cm}
\ni{\bf 5. Confronting $\E$ symmetry.} 
The Lie algebra $\lae$ is recursively defined in terms of generators
and relations (Chevalley-Serre presentation) \cite{Kac} 
\be\label{CS}
[h_i, e_j] &=& A_{ij} e_j \;,\quad [h_i, f_j] = - A_{ij} f_j \;,\quad
[e_i , f_j ] = \delta_{ij} h_j   \nonumber\\[2mm]
 && \hspace{-1.5cm} {\rm ad}(e_i)^{1- A_{ij}} (e_j) = 0 \;\; , \quad   {\rm ad} (f_i)^{1- A_{ij}} (f_j) = 0 
\ee
with ten elementary $\las$ building blocks $\{e_i, f_i ,h_i \,|\, i,j=1,...,10\}$,
that is, one for each node of the Dynkin diagram (see also \cite{GN} for a 
pedestrian introduction). Here the $\E$ Cartan matrix $A_{ij}$ 
follows from (\ref{A}) (with $r\!=\!10$), and thus has a truly (super-)gravitational 
origin \cite{DaHe01}. These `starting rules' are fairly simple to write down, but for indefinite 
Cartan matrix $A_{ij}$ 
they hide an unbounded complexity that evolves out of their repeated application
(recall  that, by contrast,  positive definite Cartan matrices lead to finite-dimensional 
simple Lie algebras, as the iterated commutators terminate after finitely many steps \cite{Kac}). 
For indefinite Cartan matrix there is thus no `closed form'  
of this Lie algebra, nor even a list or enumeration of its basis elements.
To visualize the complications it is perhaps simplest to think of it as 
a {\em Lie algebra analog of a Mandelbrot set}, 
but (for all we know) without the self-similar features. The indefiniteness of $A_{ij}$
entails the existence of infinitely many roots, and an {\em exponential increase} 
in the multiplicities of timelike imaginary roots with their length \cite{Kac}.
This feature is reminiscent of the exponential increase in the number of 
massive states in string theory, but here the growth is much more erratic:
unlike for string theory, there is no partition function nor any other
known number theoretic device to describe that growth.

To better understand the Lie algebra $\lae$ one can expand it
as a graded direct sum of vector spaces \cite{FF}
\be\label{level}
\lae \,=\, \bigoplus_{\ell = -\infty}^{\infty} {\mathfrak{e}}_{10}^{(\ell)}
\ee
where the level $\ell$ refers to a decomposition in terms of representations 
of some (preferably finite-dimensional) regular subalgebra  
${\mathfrak{e}}_{10}^{(0)}\subset\lae$.  However, any recursive analysis 
is prohibitive:  the exponential growth  of the dimensions of  
${\mathfrak{e}}_{10}^{(\ell)}$ with increasing $\ell$ is clearly evident
in the tables of \cite{FN}. No matter  where one cuts off the 
expansion (\ref{level}), this mathematical structure never reaches
a stationary state as $\ell\rightarrow\infty$ , but keeps getting more complicated! 

Exploring the physics of the $\E/\K$ model one hits upon the very same difficulties that 
obstruct the mathematical analysis (and more). The nascent complexity of this system is already
apparent in the occurrence of chaotic oscillations in the original BKL analysis \cite{BKL}.
These difficulties multiply as one probes degrees of freedom beyond the CSA.
Nevertheless, with only the Iwasawa gauge to rely on, exploiting local  $\K$ symmetry 
to restrict the degrees of freedom to non-negative levels in (\ref{level}), one can 
show that the different maximal supergravities with their duality symmetries
all emerge at low levels, depending 
on how one `slices' $\lae$ in (\ref{level}) \cite{KN0,RW,BDN}. Still, no investigation has gone beyond 
$|\ell| \leq 4$ so far, despite many efforts. The correct physical interpretation 
of the higher level modes remains an enigma.

It is common lore that non-diagonal degrees of freedom `freeze' in the BKL limit
\cite{DaHeNi03}.  Furthermore, the wave function can be shown 
to generically vanish {\em at} the singularity in this approximation \cite{KKN}.
Nevertheless, within the full $\E/\K$ coset manifold  this simplification fails
(although the evanescence of the wave function at the singularity
may still persist). Namely, inspection of the geodesic deviation equation shows that 
the sectional curvatures decrease without limit for 
imaginary roots $\alpha$, as $\a^2\!\rightarrow\! -\infty$ \cite{DaNi05}, 
implying that geodesics on $\E/\K$ become infinitely unstable as $t\!\rightarrow\!\infty$. 
It is unknown whether quantization and inclusion of fermions (perhaps in bosonized form) 
can mitigate these instabilities, but they surely affect the `wave function of the universe'.
Irrespective of how close to the singularity and at which depth in $\lae$ one specifies 
initial conditions, there always remains an uncontrollably infinite deficit: tracing the path 
of the cosmic wave packet out of the singularity would require  a descent down
a bottomless `devil's staircase'. 

\vspace*{0.2cm}
\ni{\bf 6. Complexity and information.} Although the preceding discussion provides ample
evidence for the complexity of $\lae$, this concept, though intuitively clear, should be 
properly defined and quantified. The relevant notion here is {\em algorithmic complexity},
which can be (roughly) measured via the minimal time to perform a computation \cite{QI}
(similarly, in quantum circuits `complexity' can be defined via the minimal number of 
quantum gates required to reach a given final state from a given initial one \cite{QI1}). 
If we apply this idea to $\lae$, the `complexity' of a  higher level state associated to an 
imaginary root $\alpha$ increases without bound as $\alpha^2\rightarrow -\infty$
(like the sectional curvature).
To capture the {\em intrinsic} complexity contained in the individual imaginary root
spaces would require a vastly generalized notion of automorphicity, which is  not known.

Because the time-reversed picture of the Big Bang is the black hole singularity,
the present considerations could also provide a very different perspective on
the black hole information paradox. 
The possible relevance of the BKL analysis in this 
context was recently highlighted  in \cite{Perry}, where the preservation of information
is linked to the vanishing of the wave function at the singularity (see above).
The new aspect of the present work is that for the full $\lae$ algebra, the available 
phase space at the singularity becomes infinitely larger than the
BKL phase space of diagonal metrics. Unbounded growth of 
complexity may furthermore create a basic asymmetry between moving the cosmic 
wave packet {\em into} or moving it {\em out of} the singularity,  a distinction that 
has been repeatedly emphasized by R.~Penrose, see {\em e.g.}  \cite{Penrose2}.

\vspace*{0.2cm}
\ni{\bf 7. Physics?} 
Observational confirmation of {\em any} proposal towards resolving 
the initial  singularity is notoriously difficult. An obvious idea is to search for
hidden signatures in the CMB, for instance by identifying some peculiar 
pattern in the cosmic fluctuation spectrum. However, we have only one CMB map, not 
an ensemble of maps, and this makes it difficult to ascertain whether a perceived 
special feature is not simply a statistical fluke. Indeed it appears unlikely 
(to this author) that one will  be able to unambiguously
pin down the right theory of quantum gravity simply `by looking at the sky'.

Yet, there may be different and unexpected ways to validate the present scheme.
A surprising result is that the maximal compact subgroup $\K$ admits 
unfaithful finite-dimensional fermionic representations \cite{DKN06,dBHP,KN1,KKLN}.
Remarkably, one of these can be matched with the fermion content of the 
standard model, with three generations of quarks and leptons (including right-handed 
neutrinos) \cite{MN1}; to match the fermionic quantum numbers, $\K$ is absolutely
 essential \cite{KN2,MN1}.  This is therefore a scheme that can potentially explain
the spin-$\frac12$ content of the standard model {\em as is}, with no extra room for new
fundamental spin-$\frac12$ degrees of freedom (such as those predicted by low energy
supersymmetry), and where supersymmetry is `superseded' by 
$\E$ and $\K$ symmetry. In addition, this scheme predicts eight 
supermassive spin-$\frac32$ fermions (gravitinos) 
participating in standard model interactions.
These may explain the ultrahigh energy cosmic ray events
observed over many years \cite{MN2}, and the origin and growth of primordial
black holes \cite{MN3}. Indirectly, $\E$ and $\K$ could thus help in addressing 
two major open problems of modern astrophysics.

\vspace*{0.2cm}
\ni{\bf 8. Resum\'e.} The basic claim put forward in this Note
can be summarized very simply: even if we were able eventually to formulate a 
consistent (UV complete) unified theory of quantum gravity and validate its
low energy manifestations against observation, the emergent complexity of 
this theory may prevent us from getting to the bottom of what it predicts. 
The beginning of time could thus remain beyond the reach
of our attempts to understand it.\\[3mm]
\noindent
{\bf Acknowledgments}: I am most grateful to T. Damour, M. Henneaux, 
A. Kleinschmidt and K.A. Meissner for collaboration and inspiration over
many years, and to C. Kiefer and J.L. Lehners for helpful  comments. 
I would also like to thank the referees for clarifying comments.
This work has been supported by the European Research 
 Council (ERC) under the  European Union's Horizon 2020 research and 
 innovation programme (grant agreement No 740209).


\baselineskip12pt

\begin{thebibliography}{20}

\bibitem{KT} E.W. Kolb and M.S. Turner, {\sl The Early Universe}, Addison-Wesley (1990)

\bibitem{M} V. Mukhanov, {\sl Physical Foundations of Cosmology}, Cambridge Univ. Press (2005)

\bibitem{BEG} R. Brout, F. Englert and E. Gunzig, Annals Phys. {\bf 115} (1978) 78

\bibitem{Vilenkin1} A. Vilenkin, Phys. Lett. {\bf B117} (1982) 25

\bibitem{DeWitt} B.S. DeWitt, Phys.Rev. {\bf 160} (1967) 1195; 
             Phys. Rev. {\bf 162} (1967) 1239

\bibitem{KKN} A. Kleinschmidt, M. K\"ohn and H. Nicolai, Phys. Rev. {\bf D80} (2009) 061701

\bibitem{Kiefer0} C. Kiefer, N. Kwidzinski and D. Piontek,
        Eur. Phys. J. C {\bf 79} (2019) 686

\bibitem{HH} J.B. Hartle and S.W. Hawking, Phys. Rev. {\bf D28} (1983) 2960

\bibitem{Tang} P.C.L. Tang, {\sl The ontological status of the cosmological singularity},
Springer Verlag (1989)
             
\bibitem{Kiefer} C.~Kiefer, {\sl Quantum gravity}, Clarendon Press, 2004

\bibitem{Bojo} M.~Bojowald, Phys. Rev. Lett. {\bf 86} (2001) 5227

\bibitem{Ashtekar} A. Ashtekar and T. Pawlowski, Phys. Rev. Lett. {\bf 96} (2006) 141301

\bibitem{RV} C. Rovelli and F. Vidotto, {\sl Covariant Loop Quantum Gravity},
    Cambridge Univ. Press (2014)

\bibitem{Turok} J. Khoury, B. Ovrut, P. Steinhardt and N. Turok,
Phys. Rev. {\bf D64} (2001) 123522

\bibitem{GV} M. Gasperini and G. Veneziano, Astropart. Phys. {\bf 1} (1993) 317

\bibitem{Penrose1} R. Penrose, in Proceedings of EPAC 2006, Edinburgh, Scotland

\bibitem{MRV} J. Martin, C. Ringeval and V. Vennin, Phys. Dark Univ. {\bf 5-6} (2014) 75

\bibitem{Sch} A.N.~Schellekens, Rev. Mod. Phys. {\bf 85} (2013) 1491;
                      Int. J. Mod. Phys. {\bf A30} (2015) 1530016

\bibitem{AGV} A. Borde, A.H. Guth and A. Vilenkin, Phys. Rev. Lett. {\bf 90} (2003) 151301

\bibitem{Ellis}  G.F.R.~Ellis and R. Maartens, Class. Quant. Grav. {\bf 21} (2004) 223

\bibitem{BKL} V.~A.~Belinsky, I.~M.~Khalatnikov and E.~M.~Lifshitz,
   Adv. Phys.  {\bf 19} (1970) 525

\bibitem{DaHeNi03} T.~Damour, M.~Henneaux and H.~Nicolai, 
  Class. Quant. Grav. {\bf 20} (2003) R145    

\bibitem{Misner} C.~W.~Misner, 
   Phys.\ Rev.\  {\bf 186} (1969) 1319

\bibitem{DHJN} T. Damour, M. Henneaux, B. Julia and H. Nicolai,
      Phys. Lett. {\bf B509} (2001) 323

\bibitem{DHHS} J. Demaret, J.L.Hanquin, M. Henneaux and P. Spindel,
   Phys. Lett. {\bf B175} (1986) 129.
   
\bibitem{DaHeNi02} T.~Damour, M.~Henneaux and H.~Nicolai, 
    Phys. Rev. Lett. {\bf 89} (2002) 221601 

\bibitem{DaHe01} T.~Damour and M.~Henneaux, 
  Phys.\ Rev.\ Lett. {\bf 86} (2001) 4749       

\bibitem{G} J. Greensite, Class. Quant. Grav. {\bf 13} (1996) 1339

\bibitem{TW} P.K. Townsend and M. Wohlfarth, Class. Quant. Grav. {\bf 21} (2004) 5375

\bibitem{Ganor} J. Brown, O.J. Ganor and C. Helfgott, JHEP {\bf 0408} (2004) 063

\bibitem{DS} T. Damour and P. Spindel, Phys. Rev. {\bf D95} (2017) 126011

\bibitem{CJS} E.~Cremmer, B.~Julia and J.~Scherk, 
  Phys. Lett. {\bf B76} (1978) 409

\bibitem{Ju85} B.~Julia, in: Lectures in Applied Mathematics, Vol. 21
  (1985), AMS-SIAM, p. 335    

\bibitem{West} P.C. West, Class. Quant. Grav. {\bf 18} (2001) 4443

\bibitem{Kac} V.~G.~Kac, {\sl Infinite dimensional Lie algebras}, 3rd
  edition, Cambridge University Press (Cambridge, 1990)
  
\bibitem{GN} R.W. Gebert and H. Nicolai, {\sl $\E$ for beginners}, in
{\sl Strings and Symmetries}, Springer Lecture
Notes in Physics, eds. G. Aktas et al. (1994)

\bibitem{FF} A.J. Feingold and I.B. Frenkel, Math. Ann. {\bf 263} (1983) 87

\bibitem{FN} H. Nicolai and T. Fischbacher, in {\em Kac-Moody Lie algebras
  and related topics}, eds. N. Stanumoorthy and K.C. Misra, Contemporary Mathematics
  {\bf 343}, American Mathematical Society (2004) 
  
   
\bibitem{KN0} A. Kleinschmidt and H. Nicolai, Int. J. Mod. Phys. {\bf D15} (2006) 1619
   
\bibitem{RW} F. Riccioni and P.C. West, JHEP {\bf0707} (2007) 063
   
\bibitem{BDN} E. Bergshoeff, I. DeBaetselier and T. Nutma, JHEP {\bf 0709} (2007) 047
   
\bibitem{DaNi05} T.~Damour and H.~Nicolai, 
  Class.\ Quant.\ Grav.\  {\bf 22} (2005) 2849 

\bibitem{QI} I. Wegener, {\sl Complexity Theory}, Springer-Verlag (2005)

\bibitem{QI1} M. Nielsen and J. Chuang, {\sl Quantum Computation and Quantum Information},
     Cambride Univ. Press (2010)

\bibitem{Perry} M.J. Perry, {\sl No future in black holes}, {\tt arXiv:2106.03715}

\bibitem{Penrose2} R. Penrose, 
{\sl Why current string theory cannot resolve the gravitational singularity issue},  
talk at {\em Strings 2021}, Sao Paulo

\bibitem{DKN06}  T.~Damour, A.~Kleinschmidt and H.~Nicolai,
Phys.\ Lett. {\bf B634} (2006) 319

\bibitem{dBHP} S. de Buyl, M. Henneaux and L. Paulot, JHEP {\bf 02} (2006) 056

\bibitem{KN1} A. Kleinschmidt, H. Nicolai and A. Vigan\`o, in {\em Partition functions and
 automorphic forms}, eds. V.A. Gritsenko and V.P. Spiridonov, Springer Verlag (2020)

\bibitem{KKLN} A. Kleinschmidt, R. K\"ohl, R. Lautenbacher and H. Nicolai,
 {\tt arXiv:2102.00870[math.RT]}

\bibitem{MN1} K.A. Meissner and H. Nicolai,  Phys.Rev. {\bf D 91} (2015) 065029;
Phys. Rev. Lett. {\bf 121} (2018) 091601

\bibitem{KN2} A. Kleinschmidt and H. Nicolai, Phys. Lett. {\bf B747} (2015) 251

\bibitem{MN2} K.A. Meissner and H. Nicolai, JCAP {\bf 09} (2019) 041

\bibitem{MN3} K.A. Meissner and H. Nicolai, Phys. Rev. {\bf D102} (2020) 103008; 
         Phys.Lett. {\bf B819} (2021) 136468




  



\end{thebibliography}
\end{document}